\documentclass[12pt]{amsart}
\usepackage{amssymb}
\newcommand{\norm}[1]{ \parallel #1 \parallel}
\newcommand{\beq}{\begin{equation}}
\newcommand{\eeq}{\end{equation}}
\newcommand{\bit}{\begin{itemize}}
\newcommand{\eit}{\end{itemize}}
\newcommand{\mc}{\mathcal}
\newcommand{\mb}{\mathbb}
\newcommand{\Hil}{{\mathcal H}}

\newtheorem{defn}{Definition}[section]
\newtheorem{prop}[defn]{Proposition}

\newtheorem{lemma}[defn]{Lemma}
\newtheorem{cor}[defn]{Corollary}
\theoremstyle{remark}
\newtheorem{rem}[defn]{Remark}
\newtheorem{ex}[defn]{Example}

\newcommand{\bea}{\begin{eqnarray}}
\newcommand{\ena}{\end{eqnarray}}
\newcommand{\beano}{\begin{eqnarray*}}
\newcommand{\enano}{\end{eqnarray*}}
\newcommand{\bee}{\begin{enumerate}}
\newcommand{\ene}{\end{enumerate}}
\newcommand{\A}{{\mc A}}

\newcommand{\Af}{\A_\flat}
\newcommand{\As}{\A_\sharp}
\newcommand{\Ao}{{\mc A}_0}
\newcommand{\id}{{\mb I}}

\begin{document}
\title{Morphisms of certain Banach C*-modules}
\author{Fabio Bagarello$^1$}
\author{Camillo Trapani$^2$}\footnote{\small   Dipartimento di Matematica e Applicazioni - Facolt\`a d' Ingegneria - Universit\`a di Palermo, I-90128, Palermo, Italy, e-mail: bagarell@unipa.it\\
$^2$ Dipartimento di Scienze Fisiche ed Astronomiche, Universit\`a di Palermo, I-90123, Palermo, Italy, e-mail: trapani@unipa.it \\
2000 Mathematics Subject Classification, 46H25, 46H35 }
\maketitle
\begin{abstract} Morphisms and representations of a class of Banach C*-modules, called CQ*algebras, are considered. Together with a general method for constructing CQ*-algebras, two different ways of extending the GNS-representation are presented.
\end{abstract}
\section{Introduction and Preliminaries}
Along the theory of topological quasi*-algebras, in \cite{cq1,cq2}, we undertook the study of a particular class of Banach C*-modules, called CQ*-algebras for they provide a bridge between C*-algebras and quasi *-algebras.
In \cite{cq2}, in particular, a number of structure properties of CQ*-algebras were derived, mainly concerning the subclass of {\em *-semisimple} CQ*-algebras. For their behavior, these latter seem to be a reasonable generalization of the notion of C*-algebra in the framework of partial *-algebras.
In the most general set-up a CQ*-algebra  consists of a Banach space $\A$ with involution $*$, and two C*-algebras $\Af[\|\,\|_\flat, \flat]$ and $\A_\sharp[\|\,\|_\sharp, \sharp]$ changed one into the other by $*$.
Recently, A.Inoue and the authors have shown \cite{abt} the existence of a close link between certain CQ*-algebras (called standard HCQ*-algebras) and the Tomita-Takesaki theory (and this is, in a sense, very natural).  
This fact opens the problem of the classification of (semisimple) CQ*-algebras, which should be based on a theory of representations into families of operators. This is the main motivation of this paper where we consider (and in certain cases, reconsider) the possibility of constructing concrete realizations of abstract CQ*-algebras.

The paper is organized as follows. \\
In Section 2, we prove, via a constructive proposition, the existence of several and intimately different kind of CQ*-algebras. \\
In Section 3, the basic properties of morphisms and representations are discussed; it turns out that the natural notion that leads to a reasonable definition of representation is, in this case, that of {\em bimorphism}.

Finally in Section 4, we prove the possibility of extending the well-known Gelfand - Naimark - Segal construction (GNS) to CQ*-algebras in two different ways: in the first, the starting point is a positive {\em linear functional} satisfying certain admissibility conditions; in the second, the cornerstone of the construction is, as it is natural in the framework of partial*-algebras, a positive {\em sesquilinear form} with certain invariance properties.\\
Before going forth we give, for reader's convenience, some preliminaries. 
\begin{defn}
Let $\A$ be a right Banach module over the C*-algebra $\A_\flat$, with isometric involution $*$ and such that $\A_\flat \subset \A$. Set $\A_\sharp = (\A_\flat)^*$.  We say that $\{\A, *, \A_\flat, \flat\}$ is a CQ*-algebra if
\begin{itemize}
\item[(i)] $\A_\flat$ is dense in $\A$ with respect to its norm $\|\,\|$
\item[(ii)]$\A_o:=\A_\flat \cap \A_\sharp$ is dense in $\A_\flat$ with respect to its norm  $\|\,\|_\flat$
\item[(iii)] $(AB)^* = B^*A^*, \quad \forall A,B \in \A_o$
\item[(iv)]$\|B\|_\flat = \sup_{A \in \A, \|A\|\leq 1}\|AB\|,\quad B \in \A_\flat$. 
\end{itemize}
\label{CQ}
\end{defn}
Examples of this structure have been discussed in \cite{cq1,cq2}.

Since $*$ is isometric, the space $\A_\sharp$ is itself, as it is easily seen, a C*-algebra with respect to the involution $X^\sharp:={(X^*)^\flat}^*$ and the norm $\norm{X}_\sharp :=\norm{X^*}_\flat$.\\
A CQ*-algebra is called {\em proper} if $\A_\sharp=\A_\flat$. When also $\flat =\sharp$, we indicate a proper CQ*-algebra with the notation $({\mc A},*,\Ao)$, since * is the only relevant involution and $\Ao=\A_\sharp=\A_\flat$.

From a purely algebraic point of view, each CQ*-algebra can be considered as an example of partial *-algebras, \cite{AK}, by which we mean a vector space $\A$ with involution $A \rightarrow A ^\ast$ 
[i.e. $(A+\lambda B) ^\ast =A ^\ast +\overline{\lambda} B ^\ast$ ; $A=A^{\ast \ast}$ ] and a
subset  $\Gamma \subset\A\times\A$ such that (i)  $(A,B)\in  \Gamma$  implies $(B ^\ast ,A
^\ast )\in  \Gamma$ ; (ii) $(A,B)$ and $(A,C)\in  \Gamma$  imply  $(A,B+\lambda C)\in  \Gamma$ ;
and (iii) if $(A,B)\in  \Gamma$, then there exists an element  $AB\in \A$ and for this
multiplication (which is not supposed to be associative) the following properties hold:\\ if $(A,B)\in  \Gamma$  and
$(A,C)\in  \Gamma$  then 
$AB+AC=A(B+C)$ 
and  $(AB) ^\ast =B ^\ast A ^\ast$ .

Among all CQ*-algebras a relevant role is played by the so called *-semisimple ones. 

\begin{defn}
Let $\{\A, *, \A_\flat, \flat\}$ be a  CQ*-algebra. We denote as
${\mc S}({\mc A})$ the set of sesquilinear
forms $\Omega$ on ${\mc A} \times {\mc A}$ with the following properties:
\begin{itemize}
\item[(i)]  $\Omega (A,A) \geq 0 \: \: \forall A \in {\mc A}$;
\item[(ii)] $\Omega (AB,C) = \Omega (B,A^*C) \: \forall A \in {\mc A}, \:
\forall B,C \in
{\mc A}_\flat$;
\item [(iii)] $|\Omega (A,B)| \leq \|A\| \, \|B\|\: \: \forall A,B \in
{\mc A}$.
\end{itemize}
The CQ*-algebra $({\mc A},*,{\mc A}_\flat,\flat)$ is called *-semisimple
if $\Omega(A,A)=0, \,\, \forall \Omega \in {\mc S}({\mc A})$ implies $A=0$.
\label{Definition 2.1}
\end{defn}
The definition of *-semisimple CQ*-algebras was given in \cite{cq2} for arbitrary CQ*-algebras. There, among other things, some results on functional calculus were derived, and possible refinements of the multiplications were discussed. However,  the most interesting results have been obtained for the proper case. For instance, in the abelian case, *-semisimple CQ*-algebras have been fully described. \cite{ellepi}.

\section{Construction of CQ*-algebras}
\subsection{Constructive method and first examples}
 The next  proposition extends the constructive Proposition 3.2 of \cite{cq1}. Therein it was proved that the completion of a C*-algebra $\Ao$ with respect to a norm $\|\,\|$ weaker than the C*-one, $\|\,\|_0$, for which the involution is isometric and such that $\norm {AB} \leq \norm A \norm{B}_0 $  for each
$A,B \in  {\mc
A}_0$ is a proper CQ*-algebra with $\flat =*=\sharp$. 
\begin{prop}
  Let ${\mc A}_\flat$ be a $C^\ast $-algebra, with norm $\norm{\,}_\flat $
and involution $\flat$; let $\norm{\,}$
 be another norm on $\mc A_\flat$, weaker than $\norm{\,}_\flat$
 and such that
\begin{itemize}
\item[(i)]  $\norm {AB} \leq \norm A \norm{B}_\flat \hspace{1cm}   \forall
A,B \in  {\mc
A}_\flat$ \item[(ii)]there exists a $\norm{\,}_\flat$-dense subalgebra
${\mc A}_0$ of
${\mc A}_\flat$   where an involution $^*$ (which makes of ${\mc A}_0$ a *-algebra) is defined with the property $$
\norm{A^*}=\norm{A}, \quad \forall A \in {\mc A}_0$$ \end{itemize}
then, if $\mc A$ denotes the $\norm{\,}$-completion of ${\mc A}_\flat$,
$(\mc A, *,
{\mc A}_\flat, \flat)$ is  a $CQ ^\ast $-algebra.
\label{constr}
\end{prop}

\begin{proof} Since ${\mc A}_0$ is $\norm{\,}_\flat$-dense in
${\mc
A}_\flat$, then the $\norm{\,}$-completions of ${\mc A}_0$ and ${\mc
A}_\flat$
can be identified with the same topological quasi *-algebra $\mc A$ which is
also a Banach space. Now, for $X \in {\mc A}_\flat$, put \begin{equation}
 \norm{X}_{\tilde\flat} = \sup_{\norm A \leq 1} \norm{AX}.
\label{tilde}
\end{equation}
Because of (i) we have
$$\norm{X}_{\tilde\flat} \leq \norm{X}_{\flat}, \quad \forall X \in {\mc
A}_\flat.$$
To show the converse inequality, we recall that as a consequence of Eqn.
\eqref{tilde},
${\mc A}_\flat$  with $\norm{\,}_{\tilde\flat}$
 is a normed algebra. This follows from the estimate
$$\norm {AB} \leq \norm A \norm{B}_{\tilde\flat} \hspace{1cm}   \forall
A\in \A \,,B \in  {\mc
A}_\flat.$$
 Let $X=X ^\flat \in {\mc
A}_\flat$ and let $M(X)$ denote  the abelian $C^\ast $-algebra generated by
$X$. Since
every norm that  makes an abelian $C^\ast $-algebra into a normed algebra
is necessarily
stronger than the $C^\ast $-norm \cite[Theorem 1.2.4]{sakai},  we get  the
equality
$\norm{X}_{\tilde\flat} = \norm{X}_{\flat}, \quad \forall X=X ^\flat \in
{\mc A}_\flat$.
For an arbitrary element $Y\in {\mc A}_\flat$ we  have
$$
\norm{Y}_\flat ^2 = \norm {Y ^\flat Y}_\flat=\norm {Y ^\flat
Y}_{\tilde\flat} \leq
\norm{Y^\flat}_{\tilde\flat}\norm{Y}_{\tilde\flat} $$
But $\norm{Y^\flat}_{\tilde\flat} \leq \norm{Y^\flat}_{\flat}$ and so
$\norm{Y}_\flat ^2 \leq \norm{Y}_{\flat}\norm{Y}_{\tilde\flat}$ and this
implies that
$\norm{Y}_{\flat} \leq \norm{Y}_{\tilde\flat}$. This concludes the proof.
\end{proof}
\begin{cor}Let $\{\A, *, \A_\flat, \flat\}$ be a right Banach C*-module and ${\mc B}_0$ any *-subalgebra of $\Af \cap \As$ which is also $\flat$-invariant. Let ${\mc M}_\flat ({\mc B}_0)$ be the closure of ${\mc B}_0$ in $\Af$ and ${\mc M}({\mc B}_0)$ the closure of ${\mc B}_0$ in $\A$. Then $({\mc M}({\mc B}_0), *, {\mc M}_\flat ({\mc B}_0), \flat)$ is a CQ*-algebra.
\end{cor}
\begin{proof} We notice that ${\mc M}_\flat ({\mc B}_0)$ is a C*-algebra, with respect to the involution $\flat$ and the norm $\|\,\|_\flat$, since ${\mc B}_0$ is an involutive algebra also with respect to $\flat$. The statement then follows from the previous Proposition.
\end{proof}
Let us now give some explicit applications of  Proposition \ref{constr}. 

\begin{ex} Let $S$ be an unbounded self-adjoint operator on a
Hilbert space $\Hil$
with domain $D(S)$ and with bounded inverse $S^{-1}\in B(\Hil)$,
$\|S^{-1}\|\leq 1$. We
define the commutant of the operator $S^{-1}$,
\begin{equation}
C(S^{-1}):=\left\{ X \in B(\Hil): XS^{-1}=S^{-1}X\right\}.
\label{cs}
\end{equation}
It is straightforward to check that $C(S^{-1})$ is a  C*-algebra (indeed, a von
Neumann algebra), being a norm closed *-subalgebra of $B(\Hil)$. Moreover,
 if
$S$ has not simple spectrum, $C(S^{-1})$ is not abelian \cite[II, Ch.VI,n.69]{akglaz}. If we define
on $C(S^{-1})$
a norm weaker than the norm in $B(\Hil)$, $\|.\|$, via
\begin{equation}
\|X\|_o:=\|S^{-1}XS^{-1}\|=\|S^{-2}X\|=\|XS^{-2}\|,
\label{normo}
\end{equation}
then Proposition \ref{constr} ensures us that
$(C[\|.\|_o],*,C(S^{-1})[\|.\|],*)$ is
a proper non abelian CQ*-algebra. Here we have called $C$ the
$\|.\|_o$-completion of
$C(S^{-1})$. The non triviality of the construction follows from the fact that $\|.\|_o$ is not equivalent to $\|.\|$, as it is easily checked.
We now prove the *-semisimplicity of
$(C[\|.\|_o],*,C(S^{-1})
[\|.\|],*)$. \\First we observe that any sesquilinear form of the following type
$$
\Omega_\varphi(A,B)=<S^{-1}AS^{-1}\varphi,S^{-1}BS^{-1}\varphi>,
$$
belongs to ${\mc S}(C)$ for any $\varphi\in \Hil$ with $\|\varphi\|\leq 1$.
Therefore, if $\Omega(A,A)=0$ for all $\Omega \in {\mc S}(C)$ it follows that, in particular,
$\Omega_\varphi(A,A)=0 \: \forall \varphi \in \Hil$. This implies that
$S^{-1}AS^{-1}=0$ and, therefore, that $A=0$.
\end{ex}
%\begin{ex}
% We now give a proper example of a CQ*-algebra with two
%different
%involutions following the previous Proposition. In this way we 
%generalize
% the previous example.
%Let $\Hil$ be an Hilbert space and $B(\Hil)$ the C*-algebra of the bounded
%operators on $\Hil$, with involution $\flat$ and norm $\|.\|$. Let $S$ be an
%unbounded self-adjoint operator on $\Hil$ and let us consider the spectral
%decomposition of $S^{-1}=\int_{\mc E}\lambda dE_\lambda$. We require that
%$\|S^{-1}\|\leq 1$. As before, we define
%$$
%C(S^{-1}):=\left\{X\in B(\Hil): XS^{-1}=S^{-1}X\right\},
%$$
%which is a C*-algebra w.r.t. $\flat$ and $\|.\|$. Let now $v$ be a function
%defined
%on ${\mc E}$ which takes only the values $\pm 1$. We define a new operator
%$$
%V\equiv \int_{\mc E}v(\lambda) dE_\lambda.
%$$
%It is easy to check that $V=V^\flat=V^{-1}$ and that $V\in C(S^{-1})$.  Let us %define, for any
%$X\in C(S^{-1})$, a new norm $\|X\|_S\equiv \|S^{-2}X\|$ and a new
%involution $*$,
%$X^*=V^{-1}X^\flat V$. It is obvious that, for a generic $X\in C(S^{-1})$, %$x^\flat\neq X^\sharp$, since $C(S^{-1})$ is non abelian, in general. It is also easy %to check that $X^*\in C(S^{-1})$.
% Also, one can readily verify that all the conditions of Proposition \ref{constr} are
%satisfied, so
%that $(C[\|.\|_S],*,C(S^{-1})[\|.\|],\flat)$ is a proper CQ*-algebra, $C$
%being
%the $\|.\|_S$-completion of $C(S^{-1})$.
%\end{ex}
We now describe a CQ*-algebra arising from a triplet (scale) of Hilbert spaces generated in canonical way by the operator $S$.
\begin{ex}({\em Operators on scales of Hilbert spaces}) Let ${{\mc H}}$ be a Hilbert space with scalar product
$(.,.)$ and $S$ a selfadjoint operator, with $S\geq \id$, with dense domain
$D(S)$.
In what follows it is essential that $S$ is {\em unbounded}.\\
The subspace $D(S)$ becomes a Hilbert
space,   denoted by ${\mc H}_{+1}$, with the scalar product
\begin{equation}
(f,g)_{+1} =  (Sf,Sg)
\end{equation}
Let ${\mc H}_{-1}$ denote the  conjugate dual of ${\mc H}_{+1}$. Then
${\mc H}_{-1}$
itself is a Hilbert space.

With this construction, we get, in canonical way a \underline{scale} of
Hilbert spaces
\begin{equation}
{\mc H}_{+1} \stackrel{i}{\rightarrow}{\mc H}
\stackrel{j}{\rightarrow}{\mc H}_{{-1}}
\label{OTTO}
\end{equation}
where $i$ is the
identity map $i$ of ${\mc H}_{+1}$ into ${\mc H}$  and $j$ is the
canonical embedding of  ${\mc H}$ into ${\mc H}_{-1}$ (both these maps
are continuous and with dense range).

\vspace{3mm}
\noindent With obvious identifications,  we can read
\eqref{OTTO} as a chain of topological inclusions
$$
{\mc H}_{+1} \subset {\mc H} \subset {\mc H}_{-1}. $$

By duality $S$ has an extension from ${\mc H}$ into ${\mc H}_{-1}$ which we indicate with the same symbol.
Let ${\mc B}({\mc H}_{+1},{\mc H}_{-1})$ be the Banach space of
bounded operators from ${\mc H}_{+1}$  into ${\mc H}_{-1}$
with its natural norm $\norm{\cdot}_{{+1,-1}}$.

\noindent In ${\mc B}({\mc H}_{+1},{\mc H}_{-1})$ define an involution $
A \mapsto
A^\ast$  by:

$$
(A ^\ast f,g)=\overline{(Ag,f)}   \hspace{1cm}   \forall f,g\in {\mc H}_{+1}.
$$

Then $A ^\ast \in {\mc B}({\mc H}_{+1},{\mc H}_{-1})$ and $\norm {A
^\ast}_{+1,-1}=
\norm A _{+1,-1}
 \;\;\forall A\in {\mc B}({\mc H}_{+1},{\mc H}_{-1})$.

Let ${\mc B}({\mc H}_{+1})$ be the $C^\ast $-algebra of all bounded
operators on
${\mc H}_{+1}$.
 Its natural involution is denoted
here as $\flat$ and its C*-norm as $\norm{\cdot}_\flat$.

Furthermore, let  ${\mc B}({\mc H}_{-1})$ be the $C^\ast $-algebra of
all bounded operators on ${\mc H}_{-1}$
with involution denoted as $\sharp$ and C*-norm $\norm{\cdot}_\sharp$.

\vspace{3mm}
\noindent
Then ${\mc B}({\mc H}_{+1 })$ and ${\mc B}({\mc H}_{-1})$ are
(isomorphic to)
subspaces of  ${\mc B}({\mc H}_{+1},{\mc H}_{-1})$, and $A\in {\mc B}({\mc
H}_{+1})$ if, and only if, $A ^\ast \in {\mc B}({\mc H}_{-1})$.

\vspace{3mm}
There is a {\em distinguished} *-algebra of ${\mc B}({\mc H}_{+1},{\mc
H}_{-1})$ is
$$
{\mc B} ^{+}({\mc H}_{+1}) = \{ A\in {\mc B}({\mc H}_{+1},{\mc
H}_{-1})\; 
 : A,A ^\ast \in {\mc B}({\mc H}_{+1})\; \}
$$

Clearly, if  $A\in {\mc B}({\mc H}_{+1},{\mc H}_{-1})$ and $B \in {\mc
B}({\mc
H}_{+1})$, then $AB$ is well-defined and $AB \in {\mc B}({\mc
H}_{+1},{\mc H}_{-1})$.
Analogously, if $C \in {\mc B}({\mc H}_{-1})$, $CA$ is well-defined and
$CA \in {\mc
B}({\mc H}_{+1},{\mc H}_{-1})$.

Then ${\mc B}({\mc H}_{+1},{\mc H}_{-1})$ is a {\em right Banach module}
over  the
C*-algebra ${\mc B}({\mc H}_{+1})$.\\

Now the question arises as to whether $({\mc B}({\mc
H}_{+1},{\mc H}_{-1}), *, {\mc B}({\mc H}_{+1}), \flat)$ is a
CQ*-algebra. In contrast with the original claim in \cite{cq1} this can be proved \cite{kuct} not to be true for any choice of the operator $S$ (this also solves, in negative sense a conjecture, made in \cite{ctcef}.) 

\noindent Nevertheless, Proposition \ref{constr} provides a canonical way of
constructing a CQ*-algebra of operators acting in the given scale of
Hilbert spaces.

\noindent Indeed, since ${\mc B}^+({\mc H}_{+1}) \subset {\mc B}({\mc
H}_{+1})$,  we
may consider the largest *-subalgebra ${\mc B}_0$ of ${\mc B}^+({\mc H}_{+1})$ which is also invariant with respect to the involution $\flat$  and define ${\mc B}_c({\mc H}_{+1})$ as the C*-subalgebra of ${\mc B}({\mc
H}_{+1})$ generated by
${\mc B}_0$. The non triviality of this set is discussed in \cite{kuct}. Then the conditions of Proposition \ref{constr} are fulfilled,
by choosing the
weaker norm on ${\mc B}_c({\mc H}_{+1})$ as equal to
$\norm{\cdot}_{{+1,-1}}$.
Therefore  if we denote with ${\mc B}_c({\mc H}_{+1},{\mc H}_{-1})$ the
subspace of
${\mc B}({\mc H}_{+1},{\mc H}_{-1})$ obtained by completing ${\mc B}
_c({\mc
H}_{+1})$ with respect to the norm $\norm{\cdot}_{{+1,-1}}$ we get, in any
case, a
CQ*-algebra $({\mc B}_c({\mc H}_{+1},{\mc H}_{-1}), * ,{\mc B} _c({\mc
H}_{+1}),
\flat)$ . \\
It is worth mentioning that in \cite{kuct} the *-algebra ${\mc B}_0$ has been fully described and also a characterization of ${\mc B}_c({\mc H}_{+1},{\mc H}_{-1})$ has been given.
\end{ex}
\begin{ex}
({\em CQ*-algebras of compact operators}) The same approach can be repeated starting from any  C*-subalgebra ${\mc
Q}$ of
${\mc B}_c({\mc H}_{+1})$, since conditions i) and ii) of Proposition
\ref{constr} are
satisfied whenever the weaker norm $\|\cdot\|$ is just
$\norm{\cdot}_{{+1,-1}}$ and the
adjoint is the one in ${\mc B}({\mc H}_{+1},{\mc H}_{-1})$. In
particular, we give now
an example in which all the spaces can be explicitly identified.

We start introducing the following sets of operators
$$
{\mc A}=\left\{X\in{\mc B}({\mc H}_{+1},{\mc H}_{-1}): S^{-1}X S^{-1}
\mbox{ is
compact in } \Hil\right\},
$$
$$
{\mc A}_\flat=\left\{X\in{\mc B}({\mc H}_{+1}): S X S^{-1}
  \mbox{ is
compact in }
\Hil\right\},
$$
$$
{\mc A}_\sharp=\left\{X\in{\mc B}({\mc H}_{-1}): S^{-1}X S \mbox{ is
compact in } \Hil\right\}.
$$
These sets are non empty: for instance ${\mc A}$ contains any operator of the form $SZS$, with $Z$ compact in ${\mc H}$.
As in the previous example, we indicate with the same symbol, $S$, both the operator from
$\Hil_{+1}$ into
$\Hil$ and its extension from $\Hil$ into $\Hil_{-1}$. The sets above
coincide with the
following ones: $$ {\mc A}=\left\{X\in{\mc B}({\mc H}_{+1},{\mc
H}_{-1}): X  \mbox{ is
compact from }  \Hil_{+1} \mbox{ into }  \Hil_{-1} \right\},
$$
$$
{\mc A}_\flat=\left\{X\in{\mc B}({\mc H}_{+1}): X  \mbox{ is
compact in }
\Hil_{+1} \right\},
$$
$$
{\mc A}_\sharp=\left\{X\in{\mc B}({\mc H}_{-1}): X \mbox{ is
compact in }  \Hil_{-1} \right\}.
$$
It is easy to check that ${\mc A}_\flat$ is a C*-algebra w.r.t. the
involution $\flat$
and to the norm $\|\cdot\|_\flat=\|S\cdot S^{-1}\|$.  Analogously ${\mc
A}_\sharp$ is a
C*-algebra  w.r.t. the involution $\sharp=*\flat *$ and to the norm
$\|\cdot\|_\sharp=\|S^{-1}\cdot S\|$ while ${\mc A}$ is a Banach space
w.r.t. the
involution $*$ and to the norm $\|\cdot\|=\|S^{-1}\cdot S^{-1}\|$. The norms
$\|\cdot\|_\flat$ and $\|\cdot\|$ coincide with those defined in the Banach C*-module $({\mc B}({\mc
H}_{+1},{\mc H}_{-1}), *, {\mc B}({\mc H}_{+1}), \flat)$ and the involutions $\flat$ and $*$  are the ones defined respectively in  ${\mc
B}({\mc H}_{+1})$
and ${\mc B}({\mc H}_{+1},{\mc H}_{-1})$.

In order to prove the density conditions, let us  consider the family of projection operators $P_\varphi$, $\varphi\in
\Hil_{+1}$ with
$\|\varphi\|=1$  (the norm in $\Hil$) defined by $$
P_\varphi\Psi=<\Psi,\varphi>\varphi, \quad \Psi \in \Hil.
$$
Each operator $P_\varphi$ has an obvious extension to ${\mc H}_{-1}$, which we still call
$P_\varphi$. It is straightforward to prove that $P_\varphi\in {\mc
B}({\mc H}_{+1})$,
$P_\varphi\in {\mc B}({\mc H}_{-1})$  and $P_\varphi\in {\mc A}_\flat$.
Let ${\mc
A}_0$ be the subalgebra of ${\mc A}_\flat$ generated by all the operators
$P_\varphi$, $\varphi\in \Hil_{+1}$. This is closed with respect to the adjoint $*$ and
it is also $\|\cdot\|_\flat$-dense in ${\mc A}_\flat$ since any compact operator is the
norm limit of operators of finite rank. Moreover, it is also $\|\,\|$-dense in ${\mc A}$. Applying  Proposition \ref{constr} we
conclude that
$({\mc A}[\|.\|],*{\mc A}_\flat[\|.\|_\flat],\flat)$ is a CQ*-algebra of
operators.
\end{ex}

\subsection{Constructions with families of forms}
In a previous paper, \cite{cq2}, we proposed an extension of the *-semisimplicity of the C*-algebras to CQ*-algebras, and we showed that several consequences of this definition, mainly in the field of the functional calculus, can be obtained.
Here we introduce a different definition of semisimplicity, the $\flat$-semisimplicity, that allows to define a new norm satisfying the assumptions of Proposition \ref{constr}. The application of this Proposition leads to the construction of a new CQ*-algebra whose norm closely reminds the characterization of a C*-norm in terms of states given by Gel'fand.
\begin{defn}
Let $\{\A, *, \A_\flat, \flat\}$ be a  CQ*-algebra. We denote as
${\mc S}_\flat({\mc A})$ the set of sesquilinear
forms $\Omega$ on ${\mc A} \times {\mc A}$ with the following properties:
\begin{itemize}
\item[(i)]  $\Omega (A,A) \geq 0 \: \: \forall A \in {\mc A}$;
\item[(ii)] $\Omega (AB,C) = \Omega (A,CB^\flat) \: \forall A,C \in {\mc A}, \:
\forall B \in
{\mc A}_\flat$;
\item [(iii)] $|\Omega (A,B)| \leq \|A\| \, \|B\|\: \: \forall A,B \in
{\mc A}$;
\item[(iv)] $\Omega (A,B)=\Omega (B^*,A^*), \forall A,B \in
{\mc A}$.
\end{itemize}
The CQ*-algebra $({\mc A},*,{\mc A}_\flat,\flat)$ is called $\flat$-semisimple
if $\Omega(A,A)=0, \,\, \forall \Omega \in {\mc S}_\flat({\mc A})$ implies $A=0$.
\label{Definition 6.1}
\end{defn}
It is worthwhile to remark that while conditions (i) and (iii) were already present in the definition of the family ${\mc S}({\mc A})$, condition (iv) is peculiar of this family of forms, and (ii) is a natural modification of that for ${\mc S}({\mc A})$. The non triviality of this definition is  a consequence of the results obtained in \cite{abt}, where, among other results, it is shown how the Tomita-Takesaki theory naturally provides an example of a sesquilinear form of this kind. 

In very similar way, we could speak of $\sharp$-semisimplicity, simply starting with a family of sesquilinear forms ${\mc S}_\sharp({\mc A})$, where (ii) is replaced by the specular condition
\begin{itemize}
\item[(ii$'$)]  $\Omega (AB,C) = \Omega (B,A^\sharp C) \: \forall B,C \in {\mc A}, \:
\forall A \in
{\mc A}_\sharp,$
\end{itemize}
while the other conditions are kept fixed. However, due to condition (iv) and to the equality $(X^\flat)^*=(X^*)^\sharp$, for each $X\in {\mc A}_\flat$, it is easily seen that the two families ${\mc S}_\flat({\mc A})$ and ${\mc S}_\sharp({\mc A})$ coincide. By the way, it is also interesting to remark   that without condition (iv) in the definition of the two families this equality of sets is replaced by a weaker, but still interesting, result: there is a one-to-one correspondence between forms of the two families, and this correspondence is given by the  map $\Omega  \rightarrow \Omega^*$, where:
$$
\Omega^*(A,B)\equiv \overline{\Omega(B,A)},\quad \forall A,B \in {\mc A}.
$$
It is easy to check that $\Omega\in {\mc S}_\flat({\mc A})$ iff $\Omega^*\in {\mc S}_\sharp({\mc A})$. It is evident that, either if condition (iv) is assumed or not, $\flat$-semisimplicity and $\sharp$-semisimplicity are equivalent.

One of the main reasons for the introduction of the ${\mc S}_\flat({\mc A})$ is that it allows, by means of the constructive Proposition \ref{constr}, to build up examples of CQ*-algebras starting with a given $\flat$-semisimple CQ*-algebra $\{\A, *, \A_\flat, \flat\}$. First, we introduce on $\A$ a new norm
\beq
\|A\|_{\mc S}=\sup_{\Omega\in {\mc S}_\flat({\mc A})}\Omega(A,A)^{1/2}.
\label{61}
\end{equation} 
It is not difficult to see that this is really a norm. In particular the $\flat$-semisimplicity implies that $\|A\|_{\mc S}=0$ iff $A=0$. Property (iv) of the family  ${\mc S}_\flat({\mc A})$ implies that $\|A\|_{\mc S}=\|A^*\|_{\mc S}$. Furthermore, condition (iii) implies that $\|A\|_{\mc S}\leq \|A\|$ for any $A\in \A$. In order to apply Proposition \ref{constr} we still have to check that the following inequality holds:
\beq
\|AB\|_{\mc S}\leq \|A\|_{\mc S}\|B\|,\quad  \forall A,B \in {\mc A}_\flat.
\label{62}
\end{equation}
Indeed, defining $\omega(X)=\Omega(X,\id)$ for $X\in \A_\flat$, we get $\omega(X^\flat X)\geq 0$. This implies that $\omega$ is also continuous and that the following inequality holds,
$$
|\omega(Y^\flat XY)|\leq  \omega(Y^\flat Y)\|X\|_\flat, \quad \forall X, Y \in {\mc A}_\flat.
$$
Now, inequality \eqref{62} is an immediate consequence of the definition of $\|\:\|_{\mc S}$.

Applying Proposition \ref{constr} we can conclude that $\{\A_{\mc S}, *, \A_\flat, \flat\}$ is a CQ*-algebra,
{\em containing}  $\{\A, *, \A_\flat, \flat\}$. Indeed  $\A\subset\A_{\mc S}$ since both $\A$ and $\A_{\mc S}$ are completions of the same C*-algebra $\A_\flat$ with respect to two norms, $\|\:\|$ and $\|\:\|_{\mc S}$ satisfying condition $\|\:\|_{\mc S}\leq \|\:\|$ and which are compatible in the sense of \cite{gel}; this means it is possible to extend by continuity the identity map $i: \A_\flat[\norm{\:}]\rightarrow
\A_\flat[\norm{\:}_{\mc S}]$ to their completions: $\hat i: \A[\norm{\:}]\rightarrow \A_{\mc S}[\norm{\:}_{\mc S}]$, preserving its injectivity.

Let us now prove the following

\begin{lemma}
Let $\{\A, *, \A_\flat, \flat\}$ be a $\flat$-semisimple CQ*-algebra. Then 
${\mc S}_\flat({\mc A})={\mc S}_\flat({\mc A}_{\mc S})$.
\label{LEMMA 61}
\end{lemma}

\begin{proof} Let $\Omega\in {\mc S}_\flat({\mc A}_{\mc S})$. We call $\Omega_r$ the restriction of $\Omega$ to $\A\times\A$. Since $\norm{\:}_{\mc S}$ is weaker than $\norm{\:}$, then $\Omega_r$ belongs to  
${\mc S}_\flat({\mc A})$.

Conversely, if $\Omega\in{\mc S}_\flat({\mc A})$, then, because of the following bound, $|\Omega(A,B)|\leq \Omega(A,A)^{1/2}\Omega(B,B)^{1/2}\leq 
\norm{A}_{\mc S}\norm{B}_{\mc S}$, $\Omega$ can be extended to ${\mc A}_{\mc S}\times{\mc A}_{\mc S}$ and still satisfies all the required properties.
\end{proof}

It is evident from the proof that the equality of the two sets must be understood as the possibility of associating to a form of ${\mc S}_\flat({\mc A})$ a form of ${\mc S}_\flat({\mc A}_{\mc S})$ and viceversa.

\begin{prop}
The CQ*-algebra $\{\A_{\mc S}, *, \A_\flat, \flat\}$ is $\flat$-semisimple. 
\label{Prop62}
\end{prop}

\begin{proof} Let $\Omega(X,X)=0$ $\forall \Omega\in{\mc S}_\flat({\mc A_{\mc S}})$. Then, due to the above Lemma and to the $\flat$-semisimplicity of $\{\A, *, \A_\flat, \flat\}$, we conclude that $X=0$.
\end{proof}
With the previous construction, one can always construct, starting from a $\flat$-semisimple CQ*-algebra $(\A,*,\Af,\flat)$ a new CQ*-algebra, whose norm has the form \eqref{61}, a property that closely reminds what happens for C*-algebras.

\vspace{2mm}
We now give another similar construction, starting this time from a family of sesquilinear forms on the C*-algebra $\Af$.
Let $(\A,*,\Af,\flat)$ be a CQ*-algebra. We denote with $\Sigma_\flat$ the family of all sesquilinear forms $\Omega$ on the C*-algebra $\Af$ satisfying:
\begin{itemize}
\item[(c.1)] $\Omega(A,A)\geq 0, \quad \forall A \in \Af$
\item[(c.2)] $\Omega(AB,C)=\Omega(A,CB^\flat), \quad \forall A,B,C \in \Af$
\item[(c.3)]$\Omega({\mb I},{\mb I})\leq 1$
\item[(c.4)]$\Omega(A,B)= \Omega(B^*,A^*), \quad \forall A,B \in \Af \cap \A_\sharp.$
\end{itemize}

Furthermore, assume that $(\A,*,\Af,\flat)$ satisfies the following condition:

\vspace{2mm}
\noindent [S]$ \hspace{5mm} \mbox{If }\Omega(A,A)=0, \forall \Omega \in \Sigma_\flat, \mbox{ then }A=0.$

\vspace{2mm}
\noindent Since, for each $A \in \Af$ the linear functional
$$ \omega_A(B)= \Omega(AB,A), \quad B \in \Af$$
is positive, one can easily prove the inequality
$$|\Omega(AB,A)|\leq \Omega(A,A) \|B\|_\flat, \quad \forall A,B \in \Af$$
and then
$$|\Omega(A,B)|=|\Omega(AB^\flat,\mb I)| \leq \|A\|_\flat \|B\|_\flat, \quad \forall A,B \in \Af.$$
Then we can define a new norm on $\Af$ by
$$ \|A\|_\Sigma = \sup_{\Omega \in \Sigma_\flat} \Omega(A,A)^{1/2}.$$ 

This norm defines in $\Af$ a topology coarser than that of the original norm. \\
With the help of the previous inequalities and of (c.4), we can prove that 
$$ \|AB\|_\Sigma \leq \|A\|_\Sigma \|B\|_\flat, \quad \forall A,B \in \Af $$
and
$$ \|A^*\|_\Sigma = \|A\|_\Sigma, \forall A  \in \Af \cap \A_\sharp. $$
Thus Proposition \ref{constr} applies and if $\widehat{\A}$ denotes the completion of $\Af [\|\,\|_\Sigma]$, then $(\widehat{\A}, * \Af, \flat)$ is a CQ*-algebra.\\
Now, it is easy to see that if $\Omega \in {\mc S}_\flat (\A)$ then $\Omega _{\upharpoonright \Af}$ is an element of $\Sigma_\flat$.\\
Conversely, if $\Omega_0 \in \Sigma_\flat$, since
$$ \|\Omega_0(A,B)\| \leq \|A\|_\Sigma \|B\|_\Sigma , \quad \forall A,B \in \Af $$
then, $\Omega_0$ has a continuous extension to $\widehat{\A}$ denoted as $\Omega$. It is easy to check that $\Omega \in {\mc S}_\flat (\A)$.\\
However, in spite of this very close relation between the two families of forms, $(\widehat{\A}, * \Af, \flat)$ need not be $\flat$-semisimple.\\
The interest of this construction relies on the fact that, if condition [S] holds, it is always possible to define a {\em new} CQ*-algebra where each element of $\Sigma_\flat$ is bounded.

\section{*-Isomorphisms of CQ*-algebras}
In this Section we introduce the notion of isomorphism between CQ*-algebras. We will also show that there may exist non equivalent norms which make the same C*-module into topologically different CQ*-algebras.
\begin{defn}
Let $({\mc A}, *,{\mc A}_\flat, \flat)$ and $({\mc B}, *,{\mc B}_\flat,
\flat)$ be
two CQ*-algebras. A linear map $\Phi : {\mc A} \mapsto {\mc B}$ is said
to be a
*-homomorphism of $({\mc A}, *,{\mc A}_\flat, \flat)$ into $({\mc B},
*,{\mc B}_\flat,
\flat)$ if \begin{itemize} \item[(i)] $\Phi(A^*) = \Phi(A)^*, \quad \forall
A \in {\mc
A};$ \item[(ii)]$ \Phi_\flat := \Phi \lceil_{{\mc A}_\flat}$ maps ${\mc
A}_\flat$ into
${\mc B}_\flat$ and is a *-homomorphism of C*-algebras;
\item[(iii)]$\Phi(AB)=\Phi(A)\Phi(B), \quad \forall A \in {\mc A},\, B \in
{\mc
A}_\flat$.
\end{itemize}

\noindent
A bijective *-homomorphism $\Phi$ such that $\Phi({\mc
A}_\flat)={\mc B}_\flat$ is called a *-isomorphism.
A *-homomorphism $\Phi$ is called contractive if $\norm{\Phi(A)}\leq
\norm{A}, \quad \forall A \in {\mc A}$.
A contractive *-isomorphism whose inverse is also contractive is called an
isometric
*-isomorphism. \end{defn}

\noindent {\it Remark} -- If $\Phi$ is a *-homomorphism, then
$\norm{\Phi_\flat(X)}_\flat \leq \norm{X}_\flat,\quad \forall X \in {\mc
A}_\flat$ since
each *-homomorphism of C*-algebras is contractive. Analogously, if
$\Phi_\flat$ is a
*-isomorphism, then it is necessarily isometric.

\noindent Of course, a *-homomorphism $\Phi$ can be continuous without being
contractive. \\
Finally, we notice that, if $\Phi$ is a *-isomorphism, then, by (iii),  $\Phi ({\mb I})= {\mb I}$.

\vspace{3mm}
Taking into account that a continuous isomorphism of Banach spaces has a continuous inverse, we get
\begin{prop}Let $\Phi$ be a contractive *-isomorphism of ${\mc A}$ onto
${\mc B}$. Then there exists $\gamma, \; 0<\gamma \leq 1$ such that  $$ \gamma
\norm{A} \leq \norm{\Phi(A)} \leq \norm{A}, \quad \forall A \in {\mc A}.$$
\end{prop}

\begin{prop}
Let $\Phi$ be a *-isomorphisms of $({\mc A}, *,{\mc A}_\flat, \flat)$ onto
$({\mc B}, *,{\mc B}_\flat, \flat)$. Then it is possible to define a new norm
$\norm{\,}_\Phi$ on ${\mc B}$ such that $({\mc B}, *,{\mc B}_\flat, \flat)$
is still a CQ*-algebra and $\Phi$ is isometric. \label{main} \end{prop}

\begin{proof} We define
$$\norm{B}_\Phi= \norm{\Phi^{-1}(B)}, \quad B \in {\mc B}.$$
It is very easy to prove that
\begin{itemize}
\item[(i)]${\mc B}[\norm{\,}_\Phi]$ is a Banach space;
\item[(ii)]$\norm{B^*}_\Phi = \norm{B}_\Phi, \quad \forall B \in {\mc B}$;
\item[(iii)]${\mc B}_\flat$ is $\norm{\,}_\Phi$-dense in ${\mc B}$.
\end{itemize}
Let us now define the new norm $\norm{X}^\Phi_\flat:=\sup_{\norm{A}_\Phi\leq 1 }
\norm{AX}_\Phi .$ We  prove that $\norm{X}_\flat =\norm{X}^\Phi_\flat, \quad \forall X \in {\mc B}_\flat$.

\noindent Since $\Phi^{-1}$ is necessarily an isometry between $\Af$ and ${\mc B}_\flat$, for $A \in {\mc B}$ and $X \in {\mc B}_\flat$ we have
\begin{eqnarray*}
\norm{AX}_\Phi &=&
\norm{\Phi^{-1}(AX)}\leq\norm{\Phi^{-1}(A)}\norm{\Phi^{-1}(X)}_\flat\leq \\ &\leq&
\norm{A}_\Phi \norm{X}_\flat.
\end{eqnarray*}
Therefore $\norm{X}^\Phi_\flat \leq \norm{X}_\flat, \quad \forall X \in
{\mc B}_\flat$. This inequality together with the inverse mapping theorem
already imply that the two norms are equivalent. To show that they are
exactly equal we can proceed as in the proof of Proposition \ref{constr}
after checking that $\norm{\, }^\Phi_\flat$ makes of ${\mc B}_\flat$ a
normed algebra; i.e. we need first the inequality $\norm{XY}^\Phi_\flat \leq
\norm{X}^\Phi_\flat \norm{Y}^\Phi_\flat, \quad \forall X, Y \in {\mc
B}_\flat$. But this can be easily derived from the definition of $\norm{\,
}^\Phi_\flat$ itself. The equality $\norm{X^\flat}^\Phi_\flat = \norm{X}^\Phi_\flat, \quad \forall X \in
{\mc B}_\flat$ is easy to prove.
\end{proof}

>From the previous proof one can also deduce the following
\begin{prop} Let $({\mc A}, *,{\mc A}_\flat, \flat)$ be a CQ*-algebra and
$({\mc B}, *,{\mc B}_\flat, \flat)$ a right module on ${\mc
B}_\flat$ with involution and such that ${\mc
B}_\flat\subset{\mc B}$. Let $\Phi$ be an injective linear map from ${\mc
A}$ into ${\mc B}$ with the properties: \begin{itemize} \item[(i)] $\Phi(A^*) =
\Phi(A)^*, \quad \forall A \in {\mc A};$ \item[(ii)]$ \Phi_\flat := \Phi
\lceil_{{\mc A}_\flat}$ maps ${\mc A}_\flat$ into ${\mc B}_\flat$ and is a
*-homomorphism of C*-algebras; \item[(iii)]$\Phi(AB)=\Phi(A)\Phi(B), \quad
\forall A \in {\mc A},\, B \in {\mc A}_\flat$. \end{itemize} Let us define a
norm $\norm{\, }_\Phi$ on $\Phi({\mc A})$ by the equation: $$\norm{B}_\Phi=
\norm{\Phi^{-1}(B)}, \quad B \in \Phi({\mc A}).$$ Then $(\Phi({\mc
A})[\norm{\, }_\Phi], *,{\mc B}_\flat, \flat)$ is a CQ*-algebra. \end{prop}

Now the following question arises: are there non equivalent norms on a rigged
quasi *-algebra which produce different CQ*-algebras on the same C*-algebra? The following explicit construction shows that the answer is positive. In
particular, we will give a general strategy to build up different proper CQ*-algebras 
over the
same C*-algebra and, after that, we give an explicit example.

Our starting point is a proper CQ*-algebra $({\mc A}[\|\,\|],*, {\mc A}_0[\|\,\|_0],*)$.

\begin{prop}
 Let $T$ be an unbounded, 
invertible linear map from $\mc A$ into $\mc A$ such that: $T{\mc A}_0=
{\mc A}_0$, $T(X^*)=T(X)^*$ and  $\|T(XY)\|\leq \|T(X)\| \|Y\|_0$ for all $X,Y$ in 
${\mc A}_0$. Then, defining $\|X\|_1:=\|T(X)\|$, $X\in {\mc A}_0$, 
$({\mc A}[\|\,\|_1],*, {\mc A}_0[\|\,\|_0],*)$ is a proper CQ*-algebra with norm $\|\,\|_1$ non equivalent to $\|\,\|$.
\end{prop}

\noindent {\it Remark} -- It is worth noticing 
that, in this construction, the two Banach spaces coincide while the two norms, $\|\,\|_1$
and $\|\,\|$ differ for all those elements belonging to $\mc A$ but not to $\mc A_0$, and for this reason they are not equivalent. 
This result is  a consequence of the constructive proposition for proper CQ*-algebras, see
\cite{cq1}. In particular, it is evident that the $\|\,\|_1$-completion of ${\mc A}_0$ is
exactly $\mc A$ since, on ${\mc A}_0$, $\|\,\|_1$
and $\|\,\|$ coincide.

\vspace{3mm}

An explicit example of operator $T$ can be constructed by means of the Hamel basis 
of the Banach spaces $\mc A$ and ${\mc A}_0$. \\
Indeed, let ${e_\alpha}_{\{\alpha\in J\}}$ be a Hamel basis for  $\mc A$ which contains a Hamel basis
for ${\mc A}_0$, ${e_\alpha}_{\{\alpha\in I\}}$. We can always choose $e_\alpha$ such that $e_\alpha=e_\alpha^*$.
In order to simplify things, we suppose that the set $J$ is a subset of the positive 
reals with no upper bound. We define $T$ trough its action on the basis vectors 
$e_\alpha$: $T e_\alpha=e_\alpha$ if $\alpha\in I$ and $T e_\alpha=|\alpha|e_\alpha$ if 
$\alpha\in J\backslash I$. With this definition, it is clear that $T$ is unbounded and 
invertible. Moreover, since $T$ is the 
identity map on ${\mc A}_0$, it is also evident that  $T(X^*)=T(X)^*$ for all $X\in {\mc A}_0$ and that the inequality 
$\|T(XY)\|\leq \|T(X)\| \|Y\|_0$ holds for all $X,Y$ in 
${\mc A}_0$. However, it is also clear that the norm $\|\,\|_1$ is not equivalent to $\|\,\|$, as one can check considering the values of these norms on the basis vectors.
In this way we have constructed an operator $T$ satisfying all the properties required 
above.

\begin{prop}Let $({\mc A},*,{\mc A}_\flat,\flat)$ and $({\mc B},*,{\mc
B}_\flat,\flat)$ be CQ*-algebras. Let $\Phi_\flat$ be a *-isomorphism of
${\mc A}_\flat$ onto ${\mc B}_\flat$ such that $$\norm{\Phi_\flat(X)}_{\mc B}\leq \norm{X}_{\mc A}, \, \forall X
\in {\mc A}_\flat.$$ Then $\Phi_\flat$ can be continuously extended to a
contractive *-isomorphism $\Phi$ of ${\mc A}$ onto ${\mc B}$. 

Moreover, if $\norm{\Phi_\flat(X)}_{\mc B}= \norm{X}_{\mc A}$ for all $X
\in {\mc A}_\flat$, the extension is isometric.
\end{prop}
\begin{proof} First of all define a *-isomorphism $\Phi_\sharp$
of ${\mc A}_\sharp$ onto ${\mc B}_\sharp$ by $\Phi_\sharp(Y) =
\Phi_\flat(Y^*)^*, \quad \forall Y \in {\mc A}_\sharp$. For $A \in
{\mc A}$ there exists a sequence $\{A_n\} \subset {\mc A}_\flat$
converging to $A$. Then one defines
$$ \Phi(A)= \norm{\,}_{\mc B}\lim_{n \rightarrow \infty}\Phi_\flat(A_n)$$ and
$$ \Phi(A^*)= \norm{\,}_{\mc B}\lim_{n \rightarrow \infty}\Phi_\sharp(A_n^*).$$
It is now easy to see that $\Phi$ is a contractive *-isomorphism.

The proof in the second situation is identical.
\end{proof}

In the rest of this Section we discuss briefly the problem of *-semisimplicity of CQ*-algebras and its relation with *-isomorphisms. To begin with, let $({\mc A},*,{\mc A}_\flat,\flat)$ and $({\mc B},*,{\mc
B}_\flat,\flat)$ be CQ*-algebras and let $\Phi$ be an isometric
*-isomorphism of ${\mc A}$ onto ${\mc B}$. Using Definition \ref{Definition 2.1}
 it is easy to prove that
$$ {\mc S}({\mc B})= \{ \Omega\circ \Phi^{-1}; \; \Omega \in {\mc
S}({\mc A}) \} $$ where
$$\left( \Omega\circ \Phi^{-1}\right)(A,B) = \Omega
(\Phi^{-1}(A),\Phi^{-1}(B)), \quad \forall A,B \in {\mc B}. $$

If $\Phi$ is not isometric but only contractive, then $\{ \Omega\circ
\Phi^{-1}; \; \Omega \in {\mc S}({\mc A}) \} \subset {\mc S}({\mc B})$.
Therefore we get
\begin{prop}
Let $({\mc A},*,{\mc A}_\flat,\flat)$ and $({\mc B},*,{\mc
B}_\flat,\flat)$
be CQ*-algebras and let $\Phi$ be a contractive *-isomorphism of ${\mc A}$
onto
${\mc B}$. Then, if $({\mc A},*,{\mc A}_\flat,\flat)$ is *-semisimple,
$({\mc B},*,{\mc B}_\flat,\flat)$ is also *-semisimple. \end{prop}

\begin{cor}Let $({\mc A},*,{\mc A}_\flat,\flat)$ and $({\mc B},*, {\mc
B}_\flat,\flat)$ be CQ*-algebras with ${\mc B}\subset {\mc A}$ and ${\mc
B}_\flat \subset {\mc A}_\flat$. If ${\mc B}$ is continuously embedded in
${\mc A}$ and ${\mc A}$ is *-semisimple, then ${\mc B}$ is also *-semisimple.
\end{cor}

\begin{proof}  Follows immediately from the inclusion ${\mc
S}({\mc
B}) \subseteq {\mc S}({\mc A})$.
\end{proof}

In a previous paper \cite{cq1}, in order to construct representations, the following notion was introduced
\begin{defn}
  Let $(\A,*,\Af,\flat)$ and $({\mc B},*,{\mc B}_\flat,\flat)$ be  rigged 
quasi  $^\ast $-algebras. A {\em *-bimorphism} of $(\A,*,\Af,\flat)$ 
into $({\mc B},*,{\mc B}_\flat,\flat)$ is 
a {\em pair} $(\pi,\pi_\flat)$ of linear maps $\pi:\A \mapsto {\mc B}$ and 
$\pi _\flat: \Af \mapsto {\mc B}_\flat$ such that 
\begin{itemize}
\item[(i)] $\pi_\flat$ is a homomorphism of algebras 
with $\pi_\flat(A^\flat)=\pi_\flat(A) ^\flat \;\, A\in \Af $
\item[(ii)] $\pi(A ^\ast )=\pi(A) ^\ast \;\; \forall A\in \A $
\item[(iii)] $\pi(AB)=\pi(A)\pi_\flat(B)  \forall A\in \A, B\in \Af .$
\end{itemize}
\label{DEFINITION 4.1}
\end{defn}
In general, the restriction of $\pi$ to $\Af$ is different from $\pi_\flat$. For this reason, *-homomorphisms and *-bimorphisms are different objects. Of course any *-homomorphism defines, in trivial way, a *-bimorphism, but the converse is in general not true. 
If $(\pi,\pi_\flat)$ is a *-bimorphism then $\pi(A)=\pi(\mb{I} )\pi_\flat(A)$, for each $A \in \Af$; but, in general, $\pi(\mb I )$ is different from 
$\mb I$. Obviously, if $\pi(\mb I )= {\mb I}$, then $\pi$ is a *-homomorphism.

\begin{lemma}
 If $(\pi,\pi_\flat)$ is a *-bimorphism of $(\A,*,\Af,\flat)$ into 
$({\mc B},*,{\mc B}_\flat,\flat)$ then 
$$
\pi(BA)=\pi_\sharp(B)\pi(A) \hspace{0.4cm} \forall A\in \A, \forall B\in {\mc A}_\sharp 
$$
where $\pi_\sharp(B)=\pi_\flat(B ^\ast ) ^\ast $. Moreover $\pi_\sharp$ is a homomorphism 
of ${\mc A}_\sharp$ into ${\mc B}_\sharp$ 
preserving the involution $\sharp$ of ${\mc A}_\sharp$. 
\label{LEMMA 4.2}
\end{lemma}
The proof is straightforward. 

\begin{defn}
  A {\em *-representation} of a $CQ ^\ast $-algebra 
$(\A,*,\Af,\flat)$ in the scale of Hilbert spaces ${\mc H} _{+1}
 \subset {\mc H} \subset {\mc H} _{-1}$ is a *-
bimorphism $(\pi,\pi_\flat)$ of $(\A,*,\Af,\flat)$ into the Banach right C*-module 
$({\mc B}({\mc H} _{+1},{\mc H} _{-1}),*,{\mc B}({\mc H} 
_{+1}),\flat)$ of bounded operators in the scale. The 
representation $\pi$ is said to be {\em faithful} if $Ker\, \pi = Ker\, \pi_\flat=0$. 
\end{defn}

\section{Representations of CQ*-algebras}
In this Section we will explore the possibility of constructing a GNS-like representation for a CQ*-algebra. We will give two different constructions: the first is done starting from a *-positive linear form and appears to be more adherent to the usual GNS-construction for *-algebras; the second one is obtained starting from a positive sesquilinear form, following in this way the usual path for constructing representations of partial *-algebras. 
\subsection{GNS-like construction with linear forms} 

Let $(\mc A, *,{\mc A}_\flat, \flat)$ be a CQ*-algebra (with unit $\id$). A continuous linear functional $\omega$ on $\mc A$ is called a {\em q-state} if 
{
\begin{itemize}
\item[(i)] $\omega (X^*X)\geq 0, \quad \forall X \in {\mc A}_\flat$
\item[(ii)] $\omega (X^*AY)=\overline{\omega (Y^*A^*X)}, \quad \forall A\in {\mc A},\, \forall X,Y \in {\mc A}_\flat$
\item[(iii)] $\omega({\mathbb I})=1.$
\end{itemize}
Since $\omega$ is continuous on $\A$, it turns out that it is also continuous on $\A_\flat$ with respect to $\|\,\|_\flat$; then by condition (iii) it follows that $\omega$ is $\flat$-positive, i.e. $$\omega (X^\flat X)\geq 0, \quad \forall X \in {\mc A}_\flat$$
and thus it is a state (in the usual sense) on the C*-algebra ${\mc A}_\flat$.\\
A q-state $\omega$ is said to be {\em admissible} if
\begin{itemize}
\item[(iv)]$\forall A \in \A, \exists K_A>0:|\omega (X^*AY)|^2 \leq K_A\omega(X^{\flat}X)\omega(Y^{\flat}Y),$\\$ \forall X,Y \in {\mc A}_\flat$
\item[(v)] $\{X_n\} \subset {\mc A}_\flat$ sequence s.t.  $\lim_{n \to \infty}\omega (X^*_nX_n)=0$ and \\$ \omega ((X_n-X_m)^{\flat}(X_n-X_m))\to 0\Rightarrow\lim_{n \to \infty}\omega (X^\flat_nX_n)=0.$ 
\end{itemize}}

It is worthwhile to remark that condition (v), though being a closability property, is not a consequence of the other conditions (i)-(iv) on the continuous state $\omega$.

\begin{prop} For each admissible q-state $\omega$ on a CQ*-algebra $(\mc A, *,{\mc A}_\flat, \flat)$ there exists a triplet of Hilbert spaces
$$ {\mc H}_\flat \subset{\mc H}_0\subset {\mc H}_\sharp,$$ a vector $\xi_\flat \in {\mc H}_\flat$, a linear map:
$$ \pi: \A \to {\mc B}({\mc H}_\flat ,{\mc H}_\sharp ) $$
and a $\flat$-representation 
$ \pi_\flat:  \A_\flat \to {\mc B}({\mc H}_\flat) $ of $\A_\flat$ in ${\mc H}_\flat$
such that:\\
\noindent i) $\pi (A^*)= \pi(A)^*, \quad \forall A \in \A$\\
\noindent ii) $\pi(AX)= \pi(A)\pi_\flat (X), \quad \forall A \in \A, X \in \A_\flat$\\
\noindent iii) $\omega(A)=<\pi(A)\xi_\flat,\xi_\flat>, \quad \forall A \in \A$ \\
where $<.,.>$ denotes the (extension of the) inner product of ${\mc H}_0$
\\
iv) $\xi_\flat$ is cyclic in the sense that 

 (iv.a) $\pi(\Af)\xi_\flat$ is dense in ${\mc H}_\flat[\norm{\,}_{\flat}]$,

 (iv.b) $\pi(\A)\xi_\flat$ is dense in ${\mc H}_\sharp[\norm{\,}_{\sharp}]$.\\
Moreover this representation is unique up to unitary equivalences.

\label{GNS1}
\end{prop}
\begin{proof} Since $\omega$ is $\flat$-positive on $\A_\flat$, then there exist a Hilbert space ${\mc H}_\flat$ with scalar product $<,>_\flat$ and a cyclic representation $\pi_\flat$, with cyclic vector $\xi_\flat$, such that
\beq
\omega (X)= <\pi_\flat(X)\xi_\flat,\xi_\flat>_\flat, \quad \forall X \in \A_\flat. 
\end{equation} 
It is worth recalling how $\pi_\flat$ is defined. One begins with considering the set $$ {\mc K}= \{X \in \A_\flat: \omega (X^\flat X)=0 \}$$ which is a left ideal of $\Af$. Then $\Af /{\mc K}$ is a pre-Hilbert space (the class corresponding to $X$ is denoted as $\lambda_\flat (X))$ with respect to the scalar product 
\beq
<\lambda_\flat (X),\lambda_\flat (Y)>_\flat = \omega(Y^\flat X), \quad X,Y \in \Af
\label{sp}
\end{equation}
 Then ${\mc H}_\flat$ is the completion of $\Af /{\mc K}$ with respect to the scalar product \eqref{sp} and $\pi_\flat$ is defined by
\beq
\pi_\flat (X)(\lambda_\flat (Y))= \lambda_\flat (XY), \quad X,Y \in \Af
\end{equation}
and extended by continuity to ${\mc H}_\flat$.
The cyclic vector $\xi_\flat$ is simply $\xi_\flat= \lambda_\flat (\id)$.\\
The two conditions of admissibility (iv) and (v) imply that
$$ \mc{K} = \{X \in \A_\flat: \omega (X^* X)=0\}.$$
By (iv) and (v) it follows also that ${\mc H}_\flat$ can be identified with a subspace of the completion ${\mc H}_0$ of $\Af /{\mc K}$ with respect to the scalar product 
\beq
<\lambda_\flat (X),\lambda_\flat (Y)>_0 = \omega(Y^* X), \quad X,Y \in \Af.
\label{sp2}
\end{equation}
Indeed, if $\phi \in {\mc H}_\flat$, then $\phi = \lim_{n\to \infty} \lambda_\flat( X_n)$, $X_n \in \Af$ with respect to $<,>_\flat$, so that we can associate to $\phi$ the element $\widetilde{\phi}$ of ${\mc H}_0$, which is the limit of the same sequence $\lambda_\flat( X_n)$ with respect to $<,>_0$ (by (iv) $\{ \lambda_\flat( X_n)\}$ is also a Cauchy sequence in this topology). Making use of (iv) once more we can prove that $\widetilde{\phi}$ does not depend on the particular choice of the sequence $\{ \lambda_\flat( X_n)\}$ approximating $\phi$. On the other hand (v) implies that this map is one-to-one.\\
Now, since $\omega$ is also $\sharp$-positive on $\As$, then there exist a Hilbert space ${\mc H}_\sharp$ with scalar product $<,>_\sharp$ and a cyclic representation $\pi_\sharp$, with cyclic vector $\xi_\sharp$ such that
\beq
\omega (Y)= <\pi_\sharp(Y)\xi_\sharp,\xi_\sharp>_\sharp, \quad \forall Y \in \A_\sharp 
\end{equation} 
Then we define a new representation of $\Af$ by:
$$ \widetilde{\pi}(X)= \pi_\sharp (X^{\flat *}), \quad X \in \Af.$$
The $\flat$-positivity of $\omega$ on $\Af$ implies that $\omega (X^\flat)= \overline{\omega(X)}$. Taking this fact into account we get, for each $X \in \Af$:
\begin{eqnarray*} <\widetilde{\pi}(X)\xi_\sharp,\xi_\sharp>_\sharp &=&<\pi_\sharp (X^{\flat *})\xi_\sharp,\xi_\sharp>_\sharp\\&=& \omega(X^{\flat *})= \overline{\omega(X^\flat)}=\omega(X)=<{\pi}_\flat(X)\xi_\flat,\xi_\flat>_\flat.
\end{eqnarray*}
This implies that $\widetilde{\pi}$ and $\pi_\flat$ are unitarily equivalent; i.e. there exists a unitary operator $U$, $U:{\mc H}_\flat \to {\mc H}_\sharp$ such that $U \xi_\flat =\xi_\sharp$ and $U^{-1}\widetilde{\pi}(X)U=\pi_\flat(X)$ for each $X \in \Af$.\\
By means of $U$ we may define a sesquilinear form on ${\mc H}_\flat \times {\mc H}_\sharp$ by
\beq
<\phi, F>_{\tilde 0}=<U\phi, F>_\sharp, \quad \phi \in {\mc H}_\flat, F \in {\mc H}_\sharp.
\label{form}
\end{equation}
>From \eqref{form} it follows that each element of ${\mc H}_\sharp$ can be identified with an element of the conjugate dual ${\mc H}_\flat'$ of ${\mc H}_\flat$. Indeed, if $\phi \in {\mc H}_\flat$ and  $F \in {\mc H}_\sharp$ one has:
$$|<\phi, F>_{\tilde 0}|=|<U\phi, F>_\sharp |\leq \|U\phi\|_\sharp \|F\|_\sharp = \|\phi\|_\flat \|F\|_\sharp ;$$
thus $<. ,F>_{\tilde 0}$ is continuous on ${\mc H}_\flat$.
Conversely, if $\widehat {F} \in {\mc H}_\flat'$, there exists $F \in {\mc H}_\sharp$ such that
$$<\phi, \widehat{F}>=<U\phi, F>_\sharp, \quad \phi \in {\mc H}_\flat .$$
Indeed, since $\widehat {F}$ is bounded, there exists $f \in {\mc H}_\flat$ such that
$$ <\phi, \widehat{F}>=<\phi, f>_\flat = <U\phi, U f>_\sharp = <\phi, Uf>_{\tilde 0}. $$
If we put $F=Uf$ we get the statement. An obvious consequence is that, on ${\mc H}_\flat$, $<. , .>_{0}$ coincides with $<. , .>_{\tilde 0}$.
We now define the representation $\pi$ on $\A$ by means of the following sesquilinear form on ${\mc H}_\flat \times {\mc H}_\flat$. For each $A \in \A$ we put
$$\Omega_A(\lambda_\flat (X), \lambda_\flat (Y))= \omega(Y^*AX) \quad X,Y \in \Af.$$
Condition (iv) implies the boundedness of $\Omega_A$; thus, from the previous discussion it follows that there exists a bounded operator $\pi (A)$ from ${\mc H}_\flat$ into ${\mc H}_\sharp$ such that:
\begin{equation}
 <\pi (A) \lambda_\flat (X), \lambda_\flat (Y)>_0 = \omega(Y^*AX) , \forall X,Y \in \Af.
\label{ggg}
\end{equation}
Now we have:
\begin{eqnarray*}< \pi (AZ)\lambda_\flat (X), \lambda_\flat (Y)>_0 &=& \omega(Y^*(AZ)X)=\omega(Y^*A(ZX))\\&=& < \pi (A)\lambda_\flat (ZX), \lambda_\flat (Y)>_0\\ &=& <\pi(A)\pi_\flat(Z)\lambda_\flat (X), \lambda_\flat (Y)>_0 ,
\end{eqnarray*}
therefore $\pi(AZ) =\pi(A)\pi_\flat (Z)$.\\
Moreover
\begin{eqnarray*}
< \pi (A)\lambda_\flat (X), \lambda_\flat (Y)>_0 &=& \omega(Y^*AX)=\overline{\omega(X^*A^*Y)}\\&=& \overline{< \pi (A^*)\lambda_\flat (Y), \lambda_\flat (X)>_0} \\&=& < \lambda_\flat (X), \pi (A^*)\lambda_\flat (Y)>_0.
\end{eqnarray*}
Hence $\pi(A^*)=\pi(A^*)$.\\
Finally, one has: $$\omega(A)= <\pi(A)\xi_\flat,\xi_\flat>_0$$
with $\xi_\flat = \lambda_\flat (\id)$.\\
The inclusion
$$ {\mc H}_\flat \subset{\mc H}_0\subset {\mc H}_\sharp $$
is now clear. Indeed after having proven the first one, the second inclusion is only a consequence of the identification ${\mc H}_\flat'= {\mc H}_\sharp$.\\
To conclude the proof, we need only to prove the cyclicity conditions (iv). \\
(iv.a) has already been discussed.\\
In order to prove that $\pi(\A)\lambda_\flat (\mb I)$ is dense in ${\mc H}_\sharp$, we assume that there exists $\phi \in {\mc H}_\flat$ such that
$$<\pi(A)\lambda_\flat (\mb I), \phi>_0=0, \quad \forall A \in \A .$$ 
In particular if $A \in \Af$ we have $\pi(A)=\pi(\mb I)\pi_\flat(A)$.
Then we get
$$ <\pi(A)\lambda_\flat (\mb I), \phi>_0 = <\pi_\flat(A)\lambda_\flat (\mb I),\pi(\mb I) \phi>_0 =0, \quad \forall A \in \Af,$$
and by (iv.a) this implies that $\pi(\mb I)\phi=0$.
Thus,
$$<\pi(\mb I)\phi, \lambda_\flat (Y)=0, \quad \forall Y \in \Af.$$
But, from equations \eqref{sp2} and \eqref{ggg}, $\pi(\mb I)=\mb I$ and so $\phi=0$.

The statement of the uniqueness (up to unitaries) can be proven as follows.
Let $(\hat \pi,\hat \pi_\flat )$ be another representation in the triplet of Hilbert spaces
$${\mc {\hat H}}_\flat \subset{\mc {\hat H}}_0\subset {\mc {\hat H}}_\sharp,$$
with cyclic vector $\hat \xi_\flat \in {\mc {\hat H}}_\flat$, satisfying the same conditions as $(\pi, \pi_\flat).$
Then the relations
$$ V_\flat \pi_\flat (A) \xi_\flat = \hat \pi_\flat (A) \hat \xi_\flat, \quad A \in \Af$$
$$ V \pi (A) \xi_\flat = \hat \pi (A) \hat\xi_\flat, \quad A \in \A$$
define, as is easy to check, unitary operators from ${\mc H}_\flat$ onto ${\mc {\hat H}}_\flat$ and from ${\mc H}_\sharp$ onto ${\mc {\hat H}}_\sharp$, respectively.

\end{proof}
\begin{ex}
Let $(\mc A, *,{\mc A}_\flat, \flat)$ be a CQ*-algebra of operators in a scale of Hilbert spaces, as discussed in Example 2.5.
For each $\xi \in \mc H$ with $\|\xi\|=1$ and $S^{-1}\xi= \xi$, we put:
$\omega (A) =<A\xi,\xi>, \quad A \in \A$.
Then $\omega$ is an admissible q-state on $(\mc A, *,{\mc A}_\flat, \flat)$.  
\end{ex}
As a consequence of the {\em essential} uniqueness stated above we have:
\begin{cor} Let $\omega$ be an admissible q-state on the CQ*-algebra $(\A,*,\Af, \flat)$ and $\Phi$ a *-automorphism of $(\A,*,\Af, \flat)$ such that
$$ \omega(\Phi(A))= \omega(A), \quad \forall A \in \A.$$
Then there exist uniquely determined unitaries $V$ and $V_\flat$ in the triplet of Hilbert spaces of the cyclic representation $(\pi, \pi_\flat)$ constructed in Proposition \ref{GNS1} such that
$$ V\pi (A) V^{-1} = \pi(\Phi(A)), \quad \forall A \in \A$$
and
$$ V_\flat\pi_\flat (A) V_\flat^{-1} = \pi_\flat(\Phi(A)), \quad \forall A \in \Af.$$
\end{cor}
The following proposition, which is very easy to prove, states that q-admissibility is preserved by *-isomorphisms.
\begin{prop}
Let $\Phi$ be a *-isomorphism of the CQ*-algebra $(\A,*,\Af, \flat)$ onto the CQ*-algebra $({\mc B}, * {\mc B}_\flat, \flat)$ and $\omega$ an admissible q-state for $\mc B$. Them $\omega \circ \Phi$ is an admissible q-state for $\A$.
\end{prop}
\subsection{GNS-like construction with sesquilinear forms} 
As already done in \cite{cq1}, we consider here the possibility of constructing a representation of a CQ*-algebra starting from sesquilinear forms. Our aim is to give a method which should be more natural than the one proposed in \cite{cq1}, where the construction was based on two states related among them by a certain admissibility condition. In this new GNS, on the contrary, we will use only one sesquilinear form, loosing, maybe, some freedom but gaining in clarity. 

\vspace{2mm}
\noindent Let $(\mc A, *,{\mc A}_\flat, \flat)$ be a CQ*-algebra (with unit $\id$). Let us consider a sesquilinear form $\Omega$ satisfying the following conditions
\begin{itemize}
\item[(s1)] $\Omega(A,A)\geq 0, \quad \forall A \in \A$
\item[(s2)] $\Omega(YA,B)=\Omega(A,Y^\sharp B), \quad \forall Y \in \As,\, A,B \in \A$
\item[(s3)]$|\Omega(A,B)|\leq \|A\|\|B\|, \quad \forall A,B \in \A$.
\end{itemize} 
We define
$$ \omega_\sharp (Y)= \Omega (Y, \id), \quad Y \in \As.$$
Then $\omega_\sharp$ is a positive linear form on $\As$. Therefore there exists a Hilbert space ${\mc H}_\sharp$ and a cyclic $\sharp$-representation $\pi_\sharp$ of $\As$ into ${\mc B}({\mc H}_\sharp)$, with cyclic vector $\xi_\sharp$ such that:
$$\omega_\sharp (Y) = <\pi_\sharp(Y)\xi_\sharp, \xi_\sharp >_\sharp, \quad \forall Y \in \As.$$
We now define
$$ \omega_\flat (X)= \Omega (X^{\flat *}, \id), \quad X \in \Af.$$
Since
\begin{eqnarray*} \omega_\flat (X^\flat X)&=& \Omega ((X^\flat X)^{\flat *}, \id)\\&=& \Omega (X^*X^{\flat *}, \id)= \Omega (X^{\flat *}, X^{* \sharp})=\Omega (X^{\flat *}, X^{\flat *})\geq 0,
\end{eqnarray*}
the functional $\omega_\flat$ is positive on $\Af$. Then there exists a Hilbert space ${\mc H}_\flat$ and a cyclic $\flat$-representation $\pi_\flat$ of $\Af$ into ${\mc B}({\mc H}_\flat)$, with cyclic vector $\xi_\flat$, such that:
$$\omega_\flat (X) = <\pi_\flat(X)\xi_\flat, \xi_\flat >_\flat, \quad \forall X \in \Af.$$
Now put, following an analogous path as the one in the previous Section,
$$\widetilde{\pi}(X)= \pi_\sharp (X^{\flat *}), \quad X \in \Af,$$
then $\widetilde{\pi}$ is a cyclic $\flat$-representation of $\Af$ in ${\mc H}_\sharp$ and, for each $X \in \Af$, we get
$$ \omega_\flat (X) = \Omega(X^{\flat *}, \id)=<\pi_\sharp (X^{\flat *})\xi_\sharp, \xi_\sharp>_\sharp .$$
Therefore $\widetilde{\pi}$ and $\pi_\flat$ are unitarily equivalent; i.e. there exists a unitary operator $U$, $U:{\mc H}_\flat \to {\mc H}_\sharp$ such that $U \xi_\flat =\xi_\sharp$ and $U^{-1}\widetilde{\pi}(X)U=\pi_\flat(X)$ for each $X \in \Af$.\\
We may then define a sesquilinear form on ${\mc H}_\flat \times {\mc H}_\sharp$ by
\beq
<\phi, F>=<U\phi, F>_\sharp, \quad \phi \in {\mc H}_\flat, F \in {\mc H}_\sharp
\label{sform}
\end{equation}
and in the very same way as in the proof of Proposition \ref{GNS1}, we can identify ${\mc H}_\sharp$ with the conjugate dual ${\mc H}_\flat '$ of ${\mc H}_\flat $.\\
By Riesz's lemma, any $\phi \in {\mc H}_\flat$ can be identified with a functional $F_\phi \in {\mc H}_\sharp$. So we get a Hilbert space ${\mc H}_0$ which is the completion of ${\mc H}_\flat$ with respect to the scalar product 
%\marginpar{\it check!!!}
$$<\psi,\phi>_0=<\psi,F_\phi>.$$
So we have obtained the triplet 
$${\mc H}_\flat \subset {\mc H}_0 \subset {\mc H}_\sharp.$$
Assume now that the following condition holds:
$${\rm (s4)}\hspace{4mm} \forall A \in \A, \exists K_A>0 :   \Omega (AX,AX) \leq K_A \omega_\flat (X^\flat X), \quad \forall X \in \Af$$
then, to each $A \in \A$, it corresponds a sesquilinear  form $\eta_A$ on ${\mc H}_\flat \times {\mc H}_\flat $ defined by
$$ \eta_A (\lambda_\flat (X),\lambda_\flat (Y))= \Omega (AX, Y^{\flat *}), \quad X,Y \in \Af.$$
Indeed, making use of the Cauchy-Schwarz inequality, we have:
\begin{eqnarray*}
|\eta_A (\lambda_\flat (X),\lambda_\flat (Y))|&=& |\Omega (AX, Y^{\flat *})|\\
&\leq & \Omega (AX,AX) ^{1/2}\Omega(Y^{\flat *}, Y^{\flat *})^{1/2}\\ &\leq & K_A^{1/2} \omega_\flat (X^\flat X)^{1/2} \omega_\flat (Y^\flat Y)^{1/2}.
\end{eqnarray*}
Therefore $\eta_A$ determines an operator $\pi(A)$ from ${\mc H}_\flat$ into ${\mc H}_\flat '= {\mc H}_\sharp$ such that
$$ <\pi(A)\lambda_\flat (X),\lambda_\flat (Y)>= \Omega (AX, Y^{\flat *}), \quad \forall X,Y \in \Af. $$
Now the equalities
$$<\pi (AX) \lambda_\flat (Z),\lambda_\flat (Y) > = \Omega (AXZ, Y^{\flat *})=<\pi(A)\pi_\flat(X)Z,Y>, \forall Z,X \in \Af$$
imply that $\pi(AX)=\pi(A)\pi_\flat(X), \; \forall A \in \A, X \in \Af$.
In general $\pi$ is not a *-representation. But if we add the assumption:
$${\rm (s5)}\hspace{2cm} \Omega(A,B)=\Omega(B^*,A^*) , \quad \forall A,B \in \A$$ we have:
\begin{eqnarray*}
<\pi (A) \lambda_\flat (X),\lambda_\flat (Y) > &=& \Omega (AX, Y^{\flat *})\\
&=& \Omega (AX, Y^{* \sharp})= \Omega (Y^*AX, \id)\\
&=& \Omega (\id, X^*A^*Y) = \Omega (X^{*\sharp},A^*Y)\\ &=& \Omega(X^{\flat *},A^*Y)= <\lambda_\flat (X),\pi (A^*)\lambda_\flat (Y)>.
\end{eqnarray*}
And so $\pi(A^*)=\pi(A)^*$.
We have then proved the following
\begin{prop}Let $\Omega$ be a sesquilinear form on $\A$ satisfying the conditions {\rm (s1)-(s5)}. Then there exists a triplet of Hilbert spaces
$$ {\mc H}_\flat \subset{\mc H}_0\subset {\mc H}_\sharp,$$ a vector $\xi_\flat \in {\mc H}_\flat$; a linear map:
$$ \pi: \A \to {\mc B}({\mc H}_\flat ,{\mc H}_\sharp ) $$
and a $\flat$-representation 
$ \pi_\flat:  \A_\flat \to {\mc B}({\mc H}_\flat) $ of $\A_\flat$ in ${\mc H}_\flat$
such that:\\
\noindent i) $\pi (A^*)= \pi(A)^*, \quad \forall A \in \A$\\
\noindent ii) $\pi(AX)= \pi(A)\pi_\flat (X), \quad \forall A \in \A, X \in \A_\flat$\\
\noindent iii) $\Omega(AX,Y^{\flat *})=<\pi(A)\pi_\flat(X)\xi_\flat,\pi_\flat (Y)\xi_\flat>, \quad \forall A \in \A.$
\end{prop}

\begin{rem} (1) The first remark is a consequence of condition (s5) which allows to write condition (s2) in the following equivalent form: 
\begin{itemize}
\item[(s2')] $\Omega(AY,B)=\Omega(A,BY^\flat), \quad \forall Y \in \A_\flat,\, A,B \in \A,$
\end{itemize}
which will return in the next Section.

(2) In \cite{abt} we considered {\em faithful} sesquilinear forms on a CQ*-algebra satisfying conditions (s1)-(s3) and (s5) in order to construct from it a left Hilbert algebra. CQ*-algebras that allow this construction are called {\em standard}, because they provide a link between CQ*-algebras and the Tomita-Takesaki theory.\\
 Here the additional condition (s4) has been required just to permit the construction of operators representing the given CQ*-algebra. 
\end{rem}

We end this Section noticing that the two GNS-like constructions proposed in this paper are really different: it is not difficult to see that there is no immediate relation between a general admissible state and a sesquilinear form satisfying (s1)-(s5). What we can prove, for instance, is that a sesquilinear form satisfying (s1)-(s5) and $\Omega(\id,\id)=1$ defines a state $\omega(A)=\Omega(A,\id)$, $A\in \A$, satisfying conditions (ii)-(iv) of the previous Subsection. Property (i) can be proved if we further assume that $\Omega(A^*,A^\flat)\geq 0$ for all elements is $\A_\flat$, which is a property already discussed in reference \cite{abt}.

\vspace{10mm}
{\bf\underline{Aknowledgement}}
It is a pleasure to thank Prof. K.D. K{\"u}rsten for useful discussions on the equivalence of 
proper CQ*-algebras. We also would like to thank Prof. A. Inoue for valuable discussions on the *-isomorphisms of CQ*-algebras. This work has been supported by M.U.R.S.T.

\end{document}